\documentstyle[prc,aps,eqsecnum,preprint,floats,graphicx]{revtex}

\newcommand{\bra}[1]{\langle #1 |}
\newcommand{\ket}[1]{| #1 \rangle}
\newcommand{\inproduct}[2]{\langle #1 | #2 \rangle}

\newcommand{\Ham}{{\cal H}}

\tighten
\begin{document}

\draft
\preprint{UMIST/Phys/TP/97-7} 
\title{Diabatic and Adiabatic Collective Motion in a Model Pairing System}
\author{Takashi~Nakatsukasa\footnote{
Electronic address: {\tt T.Nakatsukasa@umist.ac.uk}}
and
Niels~R.~Walet\footnote{
Electronic address: {\tt Niels.Walet@umist.ac.uk}}
}
\address{Department of Physics, UMIST, P.O.Box 88,
Manchester M60 1QD, UK}

\maketitle

\bigskip
\begin{abstract}
Large amplitude collective motion is investigated
for a model pairing Hamiltonian containing an avoided level crossing.
A classical theory of collective motion
for the adiabatic limit is applied
utilising either a time-dependent mean-field theory
or a direct parametrisation of the
time-dependent Schr\"odinger equation.
A modified local harmonic equation is formulated to take
account of the Nambu-Goldstone mode.
It turns out that in some cases the system selects a diabatic path.
Requantizing the collective Hamiltonian, a reasonable agreement
with an exact calculation for the low-lying levels are obtained for both
weak and strong pairing force.
This improves on
results of the conventional Born-Oppenheimer approximation.
\end{abstract}

\narrowtext

\section{Introduction}
\label{sec: intro}

Nuclei are finite fermionic many-body systems which
support many kinds of collective motion.
While properties of high-frequency vibrations (giant resonances)
may be well reproduced in the small-amplitude limit
by the random-phase approximation (RPA),
some low-frequency vibrations exhibit a strongly anharmonic nature
that the RPA cannot describe.
Nuclear fission and shape coexistence phenomena
also have such a large amplitude nature.
In order to investigate
these kinds of large amplitude collective motion,
one would like to reduce the number of degrees of freedom to a few
judiciously chosen slow collective coordinates.
However, in nuclear systems,
this task is not trivial, since there is no obvious separation of scales.

In molecular physics, where the masses of electrons are so much smaller
than those of the atomic nuclei,
the electronic motion is normally much faster than nuclear motion.
Thus, the collective coordinates are usually functions of
nuclear coordinates and
the Born-Oppenheimer (BO) approximation works very well.
On the other hand, in nuclear physics,
since a nucleus consists of neutrons and protons
which have almost the same masses,
both the definition of collective coordinates
and the applicability of adiabatic assumption
are never obvious.

Although a large number of studies have been done
to calculate the potential energy surface
using the constrained Hartree-Fock (HF) or Hartree-Fock-Bogoliubov (HFB) theory
with a given generalised cranking (or constraint) operator,
the choice of collective coordinate (i.e., the choice of cranking operators)
has been rather arbitrary in most cases.

In this paper we shall apply a special theory \cite{KWD91}, that
is designed to determine a self-consistent cranking operator 
for adiabatic large amplitude collective motion
(ALACM).
This theory provides a method to find approximate
decoupled motion
which is confined to a few dimensional submanifold
of the configuration space,
within the framework of classical Hamiltonian dynamics.
Since most systems of practical interest are not exactly
separable,
it is important that the theory can provide a submanifold
which is approximately decoupled.
This method has been applied to
problems in nuclear physics\cite{WDK91,WKD92}
and in other fields (see references in \cite{KWD91})

In nuclear collective motion which involves level crossings,
the problem of adiabaticity versus diabaticity
has been a long standing question since
this problem was discussed by Hill and Wheeler\cite{HW53}.
During a nuclear shape change,
the diabatic process is often more favoured than
the adiabatic one\cite{BN89}. This raises the question
whether an {\em adiabatic} theory, such as ALACM, can be used to
shed some light on diabatic dynamics.
An answer to this question is one of main goals of this paper.

In nuclear phenomena,
it is well-known that the pairing (superfluidity)
influences all  low-frequency collective motion.
A well-known example is the effect on the 
moment of inertia for rotational nuclei,
which is always smaller than the rigid-body value at low spin,
which can be explained as an effect of pairing correlations.
At the same time,
the ground states of heavy nuclei with open-shell configurations
are reasonably well described by the superfluid Bardeen-Cooper-Schrieffer
BCS  wave functions
with energy gaps of about 1 MeV.
Properties of both collective and
non-collective (quasiparticle) excitations
depend on size of the energy gap.
Furthermore, it has been argued by Bertsch\cite{Ber88} that
nuclear shape change may be associated
with the hopping of nucleon pairs by means of the pairing force.
In this case, the pairing interaction and level crossings play
an essential role to determine the collective mass (``hopping mass'').

According to these considerations, the pairing interaction
should play a key role in understanding the large amplitude collective
motion in nuclei,
especially when level crossings are involved
as the shape change is taking place.
Therefore, it is important to investigate
the applicability of ALACM
for such a system with level crossings
and a pairing force.
The theory has not been applied to such systems before.

In this paper, we study
a model Hamiltonian describing a system
interacting through the pairing force.
The model has a single-particle level crossing and
multiple local minima,
and thus may be regarded as a model for
shape-coexistence phenomena.
In Sec.~\ref{sec: ALACM}, the formalism of ALACM is briefly
recapitulated.
In order to apply the ALACM to a classical Hamiltonian with
a spurious component (Nambu-Goldstone mode),
a modified version of the local harmonic equations is formulated.
The classical Hamiltonian for a pairing Hamiltonian is derived
in Sec.~\ref{sec: model},
both by using  the mean-field (in this case BCS) theory and by applying a
parametrisation which exactly conserves particle number.
The results of numerical calculations for a simple two-level system
are given in Sec.~\ref{sec: results}
and the conclusions and an outlook are summarised in
Sec.~\ref{sec: conclusions}.

\section{Brief review of ALACM}
\label{sec: ALACM}

\subsection{Local harmonic equations (LHE)}

We briefly review the theory of ALACM
(see Ref.\cite{KWD91} for a complete description).
In this section, we use a  summation convention where
the repeated appearance of the same symbols
($\alpha,\beta,\cdots; i,j,\cdots$)
in upper and lower indices
denotes a sum over that symbol for all possible values.
We also use the convention that a comma in a lower index
indicates the derivative with respect to the coordinate,
thus $F_{,\alpha} = \partial F/\partial \xi^\alpha$.

The theory of adiabatic large amplitude collective motion
(ALACM) is applicable to a classical Hamiltonian system
which has kinetic terms only quadratic in momentum.
We thus have to start with a truncated Hamiltonian
\begin{equation}
\label{H_ad}
\Ham(\xi,\pi) = \frac{1}{2} 
         B^{\alpha \beta} \pi_\alpha \pi_\beta + V(\xi) \ ,
\hspace{1cm} \alpha, \beta = 1,\cdots,n \ ,
\end{equation}
where the mass tensor $B^{\alpha \beta}$, in general, depends on
the coordinates $\xi^\alpha$ and is defined by
truncation of the Hamiltonian to second order
\begin{equation}
B^{\alpha \beta} = \left.
        \frac{\partial^2 \Ham}{\partial\pi_\alpha \partial\pi_\beta}
        \right|_{\pi=0} \ .
\end{equation}
Thus all terms more than quadratic in momentum are neglected.
In the sense that the higher-order terms are small,
this theory may be regarded as an {\it adiabatic} theory
in the small-velocity limit.
The tensor $B_{\alpha \beta}$, which is defined as the inverse of
$B^{\alpha \beta}$
($B^{\alpha \gamma} B_{\gamma \beta} = \delta^\alpha_\beta$),
plays the role of metric tensor in the Riemannian formulation of
local harmonic equations (LHE) below.

Collective coordinates $q^i$
and intrinsic (non-collective) coordinates $q^a$
which are approximately decoupled from each other,
are assumed to be obtainable by making a point transformation,
conserving the quadratic nature of Eq.(\ref{H_ad}),
\begin{eqnarray}
q^i &=& f^i(\xi)  \quad (i=1,\cdots, K) \ ,\\
q^a &=& f^a(\xi)  \quad (a=K+1,\cdots, n) \ .
\end{eqnarray}
In this section, we use symbols ($\alpha,\beta,\cdots$) for
indices of original coordinates,
($\mu,\nu,\cdots$) for new coordinates after the transformation,
($i,j,\cdots$) for collective coordinates
and ($a,b,\cdots$) for intrinsic coordinates.
The new Hamiltonian after the point transformations takes the form,
\begin{equation}
\label{H_new}
\bar{\Ham}
         \approx \frac{1}{2} \bar{B}^{ij} p_i p_j
         + \frac{1}{2} \bar{B}^{ab} p_a p_b
            + \bar{V}(q^i,q^a) \ ,
\end{equation}
where we have used the fact that for decoupled motion the mass tensor
must be block-diagonal,
\begin{equation}
\label{mass_diagonal}
\bar{B}^{ai} = 0 \ .
\end{equation}
Besides the block-diagonality of mass tensor,
we require the absence of both ``real'' and ``geometrical'' forces
orthogonal to the decoupled manifold,
\begin{eqnarray}
\label{real_force_cond}
\bar{V}_{,a} = 0 \ ,\\
\label{geometrical_force_cond}
\bar{B}^{ij}_{,a} = 0 \ .
\end{eqnarray}

In practice, these three decoupling conditions,
(\ref{mass_diagonal}),(\ref{real_force_cond}) and
(\ref{geometrical_force_cond}), cannot be
satisfied exactly, except for special cases.
Thus, we need a method applicable to an approximately decoupled manifold.
The Riemannian formulation of LHE\cite{KWD91} is the one
we choose to use in this paper.
In a case of a single collective coordinate ($K=1$),
the basic equations of this formalism can be written as
\begin{eqnarray}
\label{LHE_1}
V_{,\alpha} = \lambda f^1_{,\alpha} \ ,\\
\label{LHE_2}
B^{\beta\gamma} V_{;\alpha\gamma} f^1_{,\beta} = \omega^2 f^1_{,\alpha}\ .
\end{eqnarray}
Here the covariant derivative (denoted by $;$) in the l.h.s.
of Eq.(\ref{LHE_2}) is defined by
\begin{equation}
V_{;\alpha\beta} \equiv V_{,\alpha\beta} 
               - \Gamma^\gamma_{\alpha\beta} V_{,\gamma} \ ,
\end{equation}
where the affine connection $\Gamma$ is defined with the help of
metric tensor $B_{\alpha\beta}$ as
\begin{equation}
\label{affine_1}
\Gamma^\alpha_{\beta\gamma} = \frac{1}{2}
   B^{\alpha\delta} \left( B_{\delta\beta, \gamma}
                        + B_{\delta\gamma , \beta}
                        - B_{\beta\gamma , \delta} \right) \ .
\end{equation}

The equations (\ref{LHE_1}) and (\ref{LHE_2}) can be solved
iteratively, starting from a stationary point.
In principle, the procedure to find a collective path is
to find successive points at which an eigenvector $f^1_{,\alpha}$
of the covariant RPA equation (\ref{LHE_2}) satisfies the force condition
(\ref{LHE_1}) at the same time.

Once we get a collective path $\Sigma$
in multi-dimensional configuration space,
a collective Hamiltonian is defined by
evaluating the Hamiltonian (\ref{H_new})
on the path $\Sigma$,
\begin{equation}
\bar{\Ham}_{\rm col} = \left.\bar{\Ham}\right|_{\Sigma, \pi_a = 0}
         \approx \frac{1}{2} \bar{B}^{11} p_1^2
            + \bar{V}(q^1) \ ,
\end{equation}
where we assume a single collective coordinate $K=1$.

The quality of decoupling can be measured by
comparing two different collective
mass parameters that can be calculated in the theory.
If we calculate the derivatives
$d\xi^\alpha/dq^1$ in terms of the tangents of the path,
\begin{equation}
\bar{B}_{11} =
    \frac{d\xi^\alpha}{dq^1} B_{\alpha\beta} \frac{d\xi^\beta}{dq^1} \ .
\end{equation}
The other mass parameter can be obtained by using the eigenvectors
$f^1_{,\alpha}$ of the covariant RPA equation.
\begin{equation}
\tilde{B}^{11} = f^1_{,\alpha} B^{\alpha\beta} f^1_{,\beta} \ .
\end{equation}
This is equal to $(\bar{B}_{11})^{-1}$ if the decoupling is exact.
Therefore, we define the decoupling measure $D$ as
\begin{equation}
\label{decoupling}
D = \left(\bar{B}_{11} \right)^{-1} \tilde{B}^{11} - 1 \ .
\end{equation}
The size of this measure $D$ indicates the badness of decoupling.

If the decoupling is good, the motion orthogonal to $\Sigma$
(motion in directions of $q^a$)
becomes irrelevant in classical systems.
However, it is not necessarily the case in quantum systems,
because, according to the uncertain relation principle,
we cannot require $q_a = 0$ (the motion is confined on $\Sigma$)
and $\pi_a = 0$ at the same time.
Therefore, one may need to include
the energy correction into the potential $\bar{V}(q)$,
which arises from quantum fluctuation
with respect to the intrinsic degrees of freedom\cite{WKD92}.
Subsequently,
the collective Hamiltonian $\bar{\Ham}_{\rm col}$
will be quantised in a flat space ($\bar{B}^{11} = 1$)
to obtain physical quantities,
such as energies and wave functions.

\subsection{Constrained local harmonic equations (CLHE)}
\label{sec: CLHE}

The most practical and straightforward way to investigate a pairing
Hamiltonian would be
to utilise the mean-field (BCS) approximation
(Sec.~\ref{sec: BCS_parametrization})
in which the general product wave functions are no longer
eigenstates of the particle number.
When an intrinsic ground state breaks a (continuous) symmetry
of Hamiltonian,
generally speaking,
a Nambu-Goldstone (NG) mode will appear.
This mode corresponds to an ``excitation'' to connect degenerate vacua.
In the case of particle-number breaking, this is often called
``pairing rotation''.
We are interested in finding a collective coordinate orthogonal to
this trivial mode.

It is also well-known that, at equilibrium points,
the NG mode has zero excitation energy and
is decoupled from the other modes in the RPA order.
Thus, one may easily separate this mode from the other physical modes
(unless the physical mode happens to be zero energy).
However, the NG mode does not necessarily have zero energy
at non-equilibrium points.
Since we have to solve the LHE at each point on
the collective path,
it is not trivial to distinguish the NG mode from the other modes.
Therefore, in this section,
we provide a modified formulation of LHE which
fixes a value of the coordinate corresponding to the NG mode.
In the BCS parametrisation, we need the constraints to
fix the particle number and gauge angle
(Sec.~\ref{sec: BCS_parametrization}). We shall make use
of the fact that the coordinate and momenta corresponding to the NG
mode are explicitly known.

We shall only discuss
the case of a single collective coordinate $K=1$ and
a single NG mode.
First we divide the set $\{ q^\mu \}$ into three subsets,
$q^1$, $q^{\sc ng}$, and $q^a$, $a=3,\cdots, n$,
which are assumed to be obtained by point transformations
\begin{eqnarray}
q^\mu = f^\mu (\xi) \ , \hspace{1cm} \xi^\alpha = g^\alpha (q) \ ,\\
p_\mu = g^\alpha_{,\mu} \pi_\alpha \ , \hspace{1cm}
                         \pi_\alpha = f^\mu_{,\alpha} p_\mu \ .
\end{eqnarray}
$q^{\sc ng}$ represents a coordinate for a NG mode and
$q^1$ for a collective coordinate now we are trying to determine.
In the case that the translational symmetry is broken,
$q^{\sc ng}$ and $p_{\sc ng}$ would be the centre-of-mass coordinate
and the total momentum, respectively.
In the case of particle-number breaking,
they correspond to the particle number and the gauge angle.
In all these cases we know the explicit form of the coordinates
and momenta for the NG mode(s).
Thus, we can write $q^{\sc ng}$ and $p_{\sc ng}$ as
functions of original coordinates $\xi^\alpha$ and momenta $\pi_\alpha$.
Although in general $q^{\sc ng}$ and $p_{\sc ng}$ have arbitrary
dependence on $\xi$ and $\pi$,
we expand them with respect to $\pi$ up to the zero-th order
for $q^{\sc ng}$ and up to first order for $p_{\sc ng}$.
In keeping with the adiabatic character of the theory,
then, we get $f^{\sc ng}(\xi^\alpha)$ and
$g^\alpha_{,\sc ng}(f^\mu(\xi))$.
An example will be shown in the next section for the particle number
and the gauge angle.

The conditions for decoupling are again given by three equations
(\ref{mass_diagonal}), (\ref{real_force_cond}) and
(\ref{geometrical_force_cond}),
where $i=1$ and {\sc ng}.
In this section we use symbols ($i,j,\cdots$)
representing 1 and {\sc ng}.
In Ref.\cite{KWD91} it is shown that the third decoupling condition
(\ref{geometrical_force_cond}) implies that the decoupled surface
is a geodesic.
The geodesic surface is defined by differential equations
\begin{eqnarray}
\label{geodesic_1}
g^\alpha_{,ij} + \Gamma^\alpha_{\beta\gamma} g^\beta_{,i} g^\gamma_{,j}
               - \bar\Gamma^k_{ij} g^\alpha_{,k} &=& 0 \ ,\\
\label{geodesic_2}
f^i_{,\alpha\beta} - \Gamma^\gamma_{\alpha\beta} f^i_{,\gamma}
               + \bar\Gamma^i_{jk} f^j_{,\alpha} f^k_{,\beta} &=& 0 \ ,
\end{eqnarray}
where
the affine connection $\Gamma$ is defined by Eq.(\ref{affine_1})
and $\bar\Gamma$ in new coordinates is in the same manner
\begin{equation}
\label{affine_2}
{\bar\Gamma}^\lambda_{\mu\nu} = \frac{1}{2}
   \bar B^{\lambda\kappa} \left( \bar B_{\kappa\mu, \nu}
                        + \bar B_{\kappa\nu , \mu}
                        - \bar B_{\mu\nu , \kappa} \right) \ .
\end{equation}

Now we consider constraints
\begin{eqnarray}
q^{\sc ng} &=& f^{\sc ng}(\xi^\alpha) = q^{\sc ng}_0
              \quad(=\mbox{const.}) \ ,\\
p_{\sc ng} &=& g^\alpha_{,\sc ng} \pi_\alpha = 0 \ ,
\end{eqnarray}
which freeze the NG degree of freedom.
The Poisson bracket between these two constraints is
$\{ q^{\sc ng}, p_{\sc ng} \}_{\rm PB}
 = f^{\sc ng}_{,\alpha} g^\alpha_{,\sc ng} = 1$.
In order to facilitate the calculation of differentiation,
we need to define a Dirac bracket\cite{Dir64} by
\begin{equation}
\{ F, G \}_{\rm DB} \equiv
\{ F, G \}_{\rm PB}
 + \{ F, q^{\sc ng} \}_{\rm PB}
   \{ p_{\sc ng}, G \}_{\rm PB}
 - \{ F, p_{\sc ng} \}_{\rm PB}
   \{ q^{\sc ng}, G \}_{\rm PB} \ .
\end{equation}
First we consider the force condition corresponding to Eq.(\ref{LHE_1}).
We differentiate the potential $V(\xi) = \bar V(q)$ with respect to
$\xi^\alpha$ keeping the constraints $q^{\sc ng} = q_0^{\sc ng}$
and $p_{\sc ng} = 0$.
\begin{equation}
\left. \frac{\partial V}{\partial \xi^\alpha}
      \right|_{q^{\sc ng}=q^{\sc ng}_0} =
\bar V_{,1} f^1_{,\alpha} + \bar V_{,a} f^a_{,\alpha}
= \bar V_{,1} f^1_{,\alpha} \ ,
\end{equation}
where the decoupling condition (\ref{real_force_cond}) has been used.
This can be calculated by using the Dirac bracket as
\begin{eqnarray}
\left. \frac{\partial V}{\partial \xi^\alpha}
      \right|_{q^{\sc ng}=q^{\sc ng}_0}
 &=& - \{ \pi_\alpha , V \}_{\rm DB} \ , \nonumber\\
 &=& V_{,\alpha} - f^{\sc ng}_{,\alpha} g^\beta_{,\sc ng} V_{,\beta} \ .
\end{eqnarray}
Equating the above two equations, we obtain the force condition
\begin{equation}
\label{CLHE_1}
V_{,\alpha} = \bar V_{,i} f^i_{,\alpha} \ .
\end{equation}
In the case of pairing rotation,
$f^{\sc ng}$ may be taken as the average particle number
${\cal N} \equiv \langle \hat N \rangle$.
Then, $\bar V_{,\sc ng} = \partial V / \partial N$ can be regarded as a
chemical potential $\mu$ and Eq.(\ref{CLHE_1}) may be
rewritten in the form similar to the constrained Hartree-Fock-Bogoliubov
equation
\begin{equation}
V_{,\alpha} = \mu {\cal N}_{,\alpha}
              + \lambda f^1_{,\alpha} \ .
\end{equation}

In order to obtain the covariant RPA equation corresponding to
Eq.(\ref{LHE_2}),
we start with the contravariant derivative of the potential
\begin{equation}
\label{temp_1}
\bar V^{,1} = V^{,\beta} f^1_{,\beta} \ ,
\end{equation}
where $V^{,\alpha} \equiv B^{\alpha\beta} V_{,\beta}$ and
$\bar V^{,1} \equiv \bar B^{1\mu} \bar V_{,\mu}
= \bar B^{1i} \bar V_{,i}$.
Again we differentiate this equation with respect to $\xi^\alpha$
in keeping with $q^{\sc ng} = p_{\sc ng} = 0$.
The derivative of the l.h.s. is easily calculated as
\begin{equation}
\label{temp_2}
\left. \frac{\partial \bar V^{,1}}{\partial\xi^\alpha}
  \right|_{q^{\sc ng} = 0}
 =  \bar V^{,1}_{,1} f^1_{,\alpha} \ .
\end{equation}
The derivative of the r.h.s. is calculated in terms of the Dirac bracket.
\begin{eqnarray}
\left. \frac{\partial}{\partial \xi^\alpha} V^{,\beta} f^1_{,\beta}
  \right|_{q^{\sc ng} = 0}
 &=& - \{ \pi_\alpha , V^{,\beta} f^1_{,\beta} \}_{\rm DB} \ ,\\
 &=& \left( V^{,\beta}_{,\alpha}
     - f^{\sc ng}_{,\alpha} g^\gamma_{\sc ng} V^{,\beta}_{,\gamma} \right)
     f^1_{,\beta}
    + V^{,\beta} \left( f^1_{,\alpha\beta}
   - f^{\sc ng}_{,\alpha} g^\gamma_{\sc ng} f^1_{,\beta\gamma}
    \right) \ ,\\
\label{temp_3}
 &=& \left( V^{,\beta}_{;\alpha}
     - f^{\sc ng}_{,\alpha} g^\gamma_{\sc ng} V^{,\beta}_{;\gamma} \right)
     f^1_{,\beta}
     - \bar\Gamma^1_{i1} {\bar V}^{,i} f^1_{,\alpha} \ ,
\end{eqnarray}
where, from the second to the third lines,
we used the geodesic equation (\ref{geodesic_2}) and
the covariant derivative defined by
\begin{equation}
V^{,\beta}_{;\alpha} \equiv V^{,\beta}_{,\alpha}
       + \Gamma^{\beta}_{\alpha\gamma} V^{,\gamma} \ .
\end{equation}
From Eqs.(\ref{temp_2}) and (\ref{temp_3}),
we obtain 
\begin{equation}
\label{CLHE_2}
\left( V^{,\beta}_{;\alpha} -
  f^{\sc ng}_{,\alpha} g^\gamma_{\sc ng} V^{,\beta}_{;\gamma} \right)
  f^1_{,\beta}
 =
  \bar V^{,1}_{;1} f^1_{,\alpha} \ .
\end{equation}
Replacing $\bar V^{,1}_{;1}$ by $\omega^2$,
this becomes the covariant RPA equation to determine
eigenfrequencies $\omega$ and eigenvecotors $f^1_{,\alpha}$.
It is worth noting that any solution $f^1_{,\alpha}$
of Eq.(\ref{CLHE_2}) with non-zero eigenfrequency 
is orthogonal to the NG mode.
This may be understood as follows:
One multiplies Eq.(\ref{CLHE_2}) by $g^\alpha_{,\sc ng}$
and takes sum over $\alpha$.
Then, since the l.h.s. becomes
$\bar V^{,1}_{,\sc ng} - \bar V^{,1}_{,\sc ng} = 0$,
it results in
\begin{equation}
\omega^2 g^\alpha_{,\sc ng} f^1_{,\alpha} = 0 \ .
\end{equation}
Therefore, with $\omega\neq 0$, we have
$g^\alpha_{,\sc ng} f^1_{,\alpha} = 0$.

Since the NG mode emerges only when we break a symmetry of
Hamiltonian,
one may expect that the CLHE formalism should become the usual LHE
when the underlying symmetry is unbroken.
In the next section, we will discuss the BCS approximation to derive
a classical Hamiltonian system.
In the parametrisation of Eq.\ (\ref{mass_parameter_BCS}) and
(\ref{potential_BCS}),
a state with $\left| \xi^\alpha \right| = 1$ and $\pi=0$
corresponds to a state in a normal phase which is an eigenstate of
particle number (from Eq.(\ref{adiabatic_gap}),
one can see the pairing gap vanish in this case).
Then, from Eq.(\ref{f_NG}),
we have $f^{\sc ng}_{,\alpha}=0$.
Therefore, the CLHE (\ref{CLHE_1}) and (\ref{CLHE_2})
become equivalent to the LHE (\ref{LHE_1}) and (\ref{LHE_2}).

The equations (\ref{CLHE_1}) and (\ref{CLHE_2}) are the defining
equations of CLHE formalism.
As in the usual LHE,
we require the self-consistency between these two equations:
A solution of Eq.(\ref{CLHE_2}) should satisfies
Eq.(\ref{CLHE_1}) at the same time.
In this paper, where we concentrate on a simple model,
it is easier to
solve the constraints $q^{\sc ng} = q_0^{\sc ng}$ and
$p_{\sc ng} = 0$ explicitly and
describe everything by independent canonical variables
$(\xi^{*\alpha},\pi^*_\alpha)$ ($\alpha=1, \cdots, n-1)$.
However,
in the realistic cases
for which it is difficult to solve the constraints,
this CLHE formalism would become more useful.

\section{Transcription into classical form}
\label{sec: model}

We consider a simple model where particles coupled to a harmonic oscillator
interact through the pairing force,
first introduced by Fukui, Matsuo and Matsuyanagi\cite{FMM91}
in order to study shape mixing.
This model can be regarded as a vibrating core plus valence particles
which can move between different levels by the pairing force,
\begin{equation}
H = H_{\rm core} + H_{\rm val} \ ,
\end{equation}
where
\begin{eqnarray}
\label{H_core}
H_{\rm core} &=& \frac{1}{2} \left( p^2 + q^2 \right) \ ,\\
\label{H_val}
H_{\rm val}  &=& \sum_{\alpha=1}^{\Omega} \epsilon_\alpha(q)
  \left( c_\alpha^\dagger c_\alpha 
   + c_{\bar\alpha}^\dagger c_{\bar\alpha} \right)
        - G \hat{P}^\dagger \hat{P} \ ,
\end{eqnarray}
and we assume $G>0$.
Here $\bar\alpha$ denotes the time-reversed state of $\alpha$ and
$\hat{P}^\dagger
 = \sum_{\alpha>0} c_\alpha^\dagger c_{\bar\alpha}^\dagger$
is the pair creation operator.
The total number of levels is $2\Omega$.
The single-particle energies $\epsilon_\alpha(q)$ depend on the
``deformation'' $q$ of the core,
which induces particle-core coupling.

We need to construct a classical Hamiltonian which describes the
dynamics of the valence particles.
In order to do this, time-dependent mean-field theory,
which is known to be a form of Hamilton's equation,
is utilised in the next section.
This is of practical interest because it has applicability to
realistic systems.
For the current model one can also
adopt a direct parametrisation of the exact
wave function to construct a classical Hamiltonian
(Sec.~\ref{sec: exact_parametrization}).

\subsection{The time-dependent mean-field (TDMF) equation}
\label{sec: BCS_parametrization}

In this section we describe a canonical parametrisation of
a general product wave function, discussed in Ref.\cite{BR86}.
We start with a BCS state $\ket{\phi_0}$ which is the vacuum to
quasiparticle operators $a_\alpha$,
\begin{eqnarray}
\label{Bogoliubov_trans_1}
a_\alpha^\dagger 
         &=& u_\alpha c_\alpha^\dagger - v_\alpha c_{\bar \alpha} \ , \\
\label{Bogoliubov_trans_2}
a_{\bar \alpha}^\dagger
         &=& u_\alpha c_{\bar \alpha}^\dagger + v_\alpha c_\alpha \ .
\end{eqnarray}
According to Thouless' theorem\cite{Tho60},
any other general product wave function $\ket{z}$ which is not
orthogonal to $\ket{\phi_0}$ may be written in the form
\begin{equation}
\ket{z} = \exp \left\{ \frac{1}{2} \sum_{\alpha\beta}
                 z_{\alpha\beta} a_\alpha^\dagger a_\beta^\dagger \right\}
          \ket{\phi_0} \ ,
\end{equation}
where $z_{\alpha\beta} = -z_{\beta\alpha}$.

The time-dependent BCS equation can be obtained by taking variations
with respect to $z$ and $z^*$ of the classical action,
\begin{equation}
\label{action_BCS}
S = \int_{t_i}^{t_f} dt
     \frac{ \bra{z} i\partial_t - H \ket{z} }{ \inproduct{z}{z} } \ .
\end{equation}
We can introduce 
canonical coordinates $(\xi_{\alpha\beta}, \pi_{\alpha\beta})$ by
$\beta_{\alpha\beta} = (\xi_{\alpha\beta} + i \pi_{\alpha\beta}) / \sqrt{2} $,
for $\alpha>\beta$. For $\alpha<\beta$ antisymmetry tells us that
the same coordinates are involved, 
\begin{equation}
\beta_{\alpha\beta}=-\beta_{\beta\alpha} = (\xi_{\beta\alpha} + 
i \pi_{\beta\alpha}) / \sqrt{2}
\end{equation}
we can easily relate $\beta$ and $z$ by
\begin{equation}
\label{beta}
\beta_{\alpha\beta}
            = \left[ z ( 1 + z^\dagger z )^{-1/2} \right]_{\alpha\beta}.
\end{equation}
This means that {\it quasi}-density matrices and {\it quasi}-pairing tensors
can be expressed by
\begin{eqnarray}
\label{quasi_density_matrix}
\bar{\rho}_{\alpha\gamma} &=&
      \frac{ \bra{z} a_\gamma^\dagger a_\alpha \ket{z} }{ \inproduct{z}{z} }
      = \left[ \beta \beta^\dagger \right]_{\alpha\gamma} \ , \\
\label{quasi_pairing_tensor}
\bar{\kappa}_{\alpha\gamma} &=&
     \frac{ \bra{z} a_\gamma a_\alpha \ket{z} }{ \inproduct{z}{z} }
     = \left[ \beta (1 - \beta^\dagger \beta)^{1/2} \right]_{\alpha\gamma} \ .
\end{eqnarray}
Using Eqs.(\ref{Bogoliubov_trans_1}) and
(\ref{Bogoliubov_trans_2}),
one can relate these quantities with {\it real} density matrices
$\rho_{\alpha\gamma} = \langle c_\gamma^\dagger c_\alpha \rangle$ and
pairing tensors $\kappa_{\alpha\gamma} = \langle c_\gamma c_\alpha \rangle$.

Then, the equations of motion take the canonical form
(using an implicit matrix notation)
\begin{eqnarray}
\label{can_eq_1}
\dot{\xi} &=&   \frac{ \partial \Ham }{ \partial \pi } \ , \\
\label{can_eq_2}
\dot{\pi} &=& - \frac{ \partial \Ham }{ \partial \xi } \ ,
\end{eqnarray}
with a classical Hamiltonian
\begin{equation}
\Ham ( \xi, \pi ) = \frac{ \bra{z} H \ket{z} }{ \inproduct{z}{z} } \ .
\end{equation}

Since monopole pairing interactions in nuclei were originally introduced
in order to give a simple model of the short-range attractive force,
often only the Hartree terms are taken into account
in the mean-field calculations.
Thus, we shall neglect the exchange terms in Sec.~\ref{sec: results_BCS}.
If we neglect the exchange terms for the Hamiltonian (\ref{H_val}),
only $(\alpha\gamma) = (\alpha\bar\alpha)$ components are available and
the matrix relations can be regarded as ordinary
{\it c}-number relations.
For instance, Eqs.(\ref{quasi_density_matrix}) and
(\ref{quasi_pairing_tensor}) now take the form
\begin{equation}
\bar{\rho}_{\alpha\bar\alpha} =
     \beta_{\alpha\bar\alpha} \beta_{\alpha\bar\alpha}^* \ , \hspace{1cm}
\bar{\kappa}_{\alpha\bar\alpha} =
 \beta_{\alpha\bar\alpha}
     (1 - \beta_{\alpha\bar\alpha}^* \beta_{\alpha\bar\alpha})^{1/2} \ ,
\end{equation}
where there is no summation with respect to the indices
$(\alpha, \bar\alpha)$.

The coordinates defined above depend on the reference state
$\ket{\phi_0}$.
For practical applications, it is probably convenient to use a local
coordinate at each point on a collective path when one solves the
LHE\cite{WDK91}.
Namely, when moving to a nearby point,
one changes the reference state to the slater determinant at this new point,
making the new point the origin of coordinate system,
and solves the LHE.
However, in this paper, in order to visualise a collective path in
the configuration space,
we use a fixed  reference state and
 a global coordinate system.
Since this global coordinate system cannot describe
a state $\ket{\phi}$ which is orthogonal to
the reference state $\ket{\phi_0}$,
we need to select a reference state suitable for describing
all relevant states.

In case that we consider a system with $N=\Omega$ particles,
we may take the BCS state with
$u_\alpha = v_\alpha = 1/\sqrt{2}$ for all $\alpha$ as the $\ket{\phi_0}$
($ \bra{\phi_0} \hat{N} \ket{\phi_0} = \Omega$).
The classical Hamiltonian for valence particles (\ref{H_val}) may be
written as
\begin{eqnarray}
\Ham_{\rm val} ( \xi, \pi ) &=& \sum_{\alpha>0} \epsilon_\alpha(q)
              \left[ 1 + \sqrt{2} \xi_\alpha \left\{ 1 - \frac{1}{2}
               ( \xi_\alpha^2 + \pi_\alpha^2 ) \right\}^{1/2} \right] 
    \nonumber\\
 &&- \frac{G}{4} \left[ \left(
        \Omega - \mbox{\boldmath $\xi$}^2 - \mbox{\boldmath $\pi$}^2
         \right)^2
         + 2 \left\{ \sum_{\alpha>0} \pi_\alpha 
          \left( 1 - \frac{1}{2} (\xi_\alpha^2 + \pi_\alpha)^2 \right)^{1/2}
                   \right\}^2 \right] \ ,
\end{eqnarray}
where we used notations $\xi_\alpha \equiv \xi_{\alpha \bar\alpha}$,
$\pi_\alpha \equiv \pi_{\alpha \bar\alpha}$,
$\mbox{\boldmath $\xi$}^2 \equiv \sum_\alpha \xi_\alpha^2$,
$\mbox{\boldmath $\pi$}^2 \equiv \sum_\alpha \pi_\alpha^2$,
and neglected the exchange terms.
Expanding this Hamiltonian up to second order in momentum,
we obtain an adiabatic Hamiltonian in the form of Eq.(\ref{H_ad}) with
\begin{eqnarray}
\label{mass_parameter_BCS}
B^{\alpha\beta}(\xi) &=& 
   \left. \frac{\partial^2 \Ham_{\rm val}}{\partial\pi_\alpha \partial\pi_\beta}
    \right|_{\pi=0}
        = \left\{ 
          -\frac{\epsilon_\alpha(q)}{\sqrt{2}} \frac{\xi_\alpha}{\Xi_\alpha}
          + G ( \Omega - \mbox{\boldmath $\xi$}^2 )
          \right\} \delta_{\alpha\beta}
          - G \Xi_\alpha \Xi_\beta \ , \\
\label{potential_BCS}
V(\xi) &=& \Ham_{\rm val} ( \xi, \pi=0 ) =
    \sqrt{2} \sum_{\alpha>0} \epsilon_\alpha(q) \xi_\alpha \Xi_\alpha 
    - \frac{G}{4} \left( \Omega-\mbox{\boldmath $\xi$}^2 \right)^2 \ ,
\end{eqnarray}
where
\begin{equation}
\Xi_\alpha = \left( 1 - \frac{1}{2} \xi_\alpha^2 \right)^{1/2} \ ,
\end{equation}
The average number of particles is
\begin{eqnarray}
\label{number}
{\cal N}(\xi,\pi) = \frac{\bra{z} \hat{N} \ket{z}}{\inproduct{z}{z}}
        &=&\Omega + \sqrt{2} \sum_{\alpha>0} \xi_\alpha
                \left\{ 1 - \frac{1}{2}
               ( \xi_\alpha^2 + \pi_\alpha^2 ) \right\}^{1/2} \ ,\\
\label{adiabatic_number}
        &\approx&\Omega + \sqrt{2} \sum_{\alpha>0} \xi_\alpha \Xi_\alpha
             + O(\pi^2) \ .
\end{eqnarray}
The pairing gap is
\begin{eqnarray}
\Delta(\xi,\pi)  = G \frac{\bra{z} \hat{P} \ket{z}}{\inproduct{z}{z}}
      &=&\frac{G}{2} \left[ \Omega - \mbox{\boldmath $\xi$}^2
                           - \mbox{\boldmath $\pi$}^2
          + \sqrt{2} i \sum_{\alpha>0} \pi_\alpha
            \left\{ 1 - \frac{1}{2} (\xi_\alpha^2 + \pi_\alpha^2 ) \right\}^{1/2}
            \right] \ , \\
\label{adiabatic_gap}
     &\approx& \frac{G}{2} \left\{ \Omega - \mbox{\boldmath $\xi$}^2
          + \sqrt{2} i \sum_{\alpha>0} \pi_\alpha \Xi_\alpha \right\}
          + O(\pi^2) \ .
\end{eqnarray}
Thus, the gauge angle $\varphi$ in the adiabatic limit
can be expressed by
\begin{equation}
\label{adiabatic_gauge_angle}
\varphi = \arctan \frac{{\rm Im} \Delta}{{\rm Re} \Delta}
     \approx \frac{\sqrt{2}}{\Omega-\mbox{\boldmath $\xi$}^2}
     \sum_{\alpha>0}\pi_\alpha \Xi_\alpha  + O(\pi^3) \ .
\end{equation}

Since we consider a system with $N=\Omega$ particles,
we may regard $({\cal N} - \Omega)/2$ as $q^{\sc ng}$
and $\varphi$ as $p_{\sc ng}$ in Sec.~\ref{sec: CLHE}.
Using the adiabatic expression of the particle number
(\ref{adiabatic_number})
and the gauge angle (\ref{adiabatic_gauge_angle}),
we can identify
\begin{eqnarray}
\label{f_NG}
f^{\sc ng}(\xi) &=& \frac{{\cal N}^{\rm ad} - \Omega}{2} =
          \frac{1}{\sqrt{2}} \sum_{\alpha>0} \xi_\alpha \Xi_\alpha \ ,\\
\label{g_NG}
\frac{\partial g^\alpha}{\partial q^{\sc ng}} (\xi) &=&
        \left. \frac{\partial \varphi}{\partial \pi_\alpha}\right|_{\pi=0}
      = \frac{\sqrt{2}}{\Omega-\mbox{\boldmath $\xi$}^2} \Xi_\alpha \ .
\end{eqnarray}
One can see that the canonical relation
\begin{equation}
\left\{ \frac{{\cal N}^{\rm ad}}{2} , \varphi^{\rm ad}
                                   \right\}_{\rm PB}
 = \sum_\alpha \frac{\partial f^{\sc ng}}{\partial \xi_\alpha}
            \frac{\partial g^\alpha}{\partial q^{\sc ng}} = 1 \ ,
\end{equation}
is satisfied.

In case of $N=\Omega=2$, the conservation of particle number in the
adiabatic limit (\ref{adiabatic_number}) is simply satisfied by
\begin{equation}
\xi_1 = -\xi_2 \ , \hspace{1cm} \pi_1 = -\pi_2\ .
\end{equation}
Then, $\Ham_{\rm val}$ can be expressed as
\begin{eqnarray}
\label{H_val_1dim}
\Ham_{\rm val} &=& \frac{1}{2} B^{\rm val} \pi^2 + V(\xi) \ , \\
\label{B_1dim}
B_{\rm val} &=& -\frac{\sqrt{2}\epsilon \xi}{4 \Xi} + G(1-\xi^2) \ ,\\
V(\xi) &=& 2\sqrt{2} \epsilon \xi \Xi - G (1 - \xi^2 )^2 \ ,
\end{eqnarray}
where we assume $\epsilon \equiv \epsilon_1 = - \epsilon_2$
and
\begin{equation}
\xi \equiv \xi_1 = -\xi_2\ , \hspace{1cm}
\pi \equiv \pi_1 - \pi_2 = 2 \pi_1 \ , \hspace{1cm}
\Xi \equiv \left( 1 - \frac{1}{2} \xi^2 \right)^{1/2} \ .
\end{equation}
The potential is stationary at
\begin{eqnarray}
\xi &=& \pm 1, \\
\label{two_states}
    && \mbox{ and } -\frac{{\rm sgn}\epsilon}{\sqrt{2}} \left(
    \sqrt{1+\frac{|\epsilon|}{G}} \pm \sqrt{1-\frac{|\epsilon|}{G}}
     \right) \ , \mbox{ for } | \epsilon | < G \ .
\end{eqnarray}
At $\xi = 1$ ($-1$) two particles occupy the $\alpha=1$ ($\alpha = 2$) level.
For $\epsilon\neq 0$, one of the states corresponds to a local minimum
and the other to a local maximum.
For $| \epsilon | < G$,
the potential has another local minimum which corresponds to
a superfluid state
(the upper and lower signs in Eq.(\ref{two_states}) give the same state
in the limit of $\pi=0$).

\subsection{The time-dependent Schr\"odinger equation (TDSE)}
\label{sec: exact_parametrization}

In case of $N=\Omega=2$,
one can easily find a canonical parametrisation
for the exact time-dependent Schr\"odinger equation (TDSE).
With a notation $\ket{\alpha\bar\alpha}=c_\alpha^\dagger c_{\bar\alpha}^\dagger \ket{0}$,
normalised states can be written in general as
\begin{equation}
\label{state_2_level}
\ket{\psi(t)} = e^{i\phi(t)} \cos (\alpha(t)) \ket{1\bar 1}
              + e^{-i\phi(t)} \sin (\alpha(t)) \ket{2\bar 2} \ .
\end{equation}

An action, similar to Eq.(\ref{action_BCS}), may be defined
by
\begin{equation}
S = \int_{t_i}^{t_f} dt
     \bra{\psi(t)} i\partial_t - H \ket{\psi(t)} \ .
\end{equation}
The time-dependent Schr\"odinger equation can be given in terms of
the least action principle.
For the Hamiltonian $H_{\rm val}$,
we can introduce canonical coordinates $(\xi,\rho)$\footnote{
In order to avoid confusion with $\pi=3.14\cdots$,
in this section we use the symbol $\rho$
for the momentum conjugate to $\xi$.}
as follows.
\begin{equation}
\xi = \alpha \ , \hspace{1cm}
\rho = - 2 \phi \sin 2\alpha \ .
\end{equation}
The canonical equations of motion (\ref{can_eq_1}) and
(\ref{can_eq_1}) hold for the Hamiltonian
\begin{equation}
\Ham_{\rm val} = \bra{\psi} H_{\rm val} \ket{\psi}
               = 2 \epsilon \cos 2\xi
                 -G \left\{ 1 + \sin 2\xi
                      \cos\left(-\frac{\rho}{\sin 2\xi} \right)\right\}
                   \ ,
\end{equation}
where $\epsilon \equiv \epsilon_1 = -\epsilon_2$.
The adiabatic Hamiltonian (\ref{H_val_1dim})
is obtained by expanding up to second order in momentum.
The resultant mass parameter and potential are
\begin{eqnarray}
\label{B_exact_param}
B_{\rm val}(\xi) &=& \frac{G}{\sin 2\xi} \ , \\
V(\xi) &=& 2\epsilon \cos 2\xi -G \left( 1 + \sin 2\xi \right) \ .
\end{eqnarray}
In this parametrisation, the mass does not depend on the single-particle
energy $\epsilon$.
It is apparent from Eq.(\ref{state_2_level}) that
$\xi = 0$ ($\pi/2$) corresponds to a pure configuration $\ket{1\bar 1}$
($\ket{2\bar 2}$).
There are two stationary points of the potential in a region
$0 \leq \xi \leq \pi$ which are specified by a condition
\begin{equation}
\label{local_minima_2_level}
\tan 2\xi = \tan 2\alpha = - \frac{G}{2\epsilon} \ .
\end{equation}
These give eigenstates (\ref{state_2_level}) of $H_{\rm val}$
with $\phi=0$.
For $G \geq 0$, the state in a region $0 \leq \xi \leq \pi/2$
always has lower energy and
the mass parameter $B_{\rm val}$ is non-negative in this region.

\section{Results}
\label{sec: results}

\begin{figure}
\centerline{\includegraphics[width=7cm]{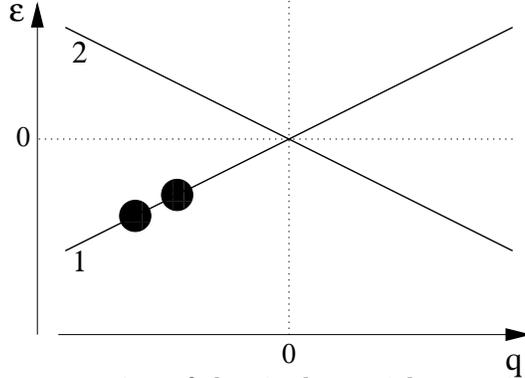}}
\caption{
A schematic representation of the single-particle energies
as a function of the coordinate $q$,
for the case $\epsilon_0 = 0$, $\chi>0$.
}
\label{Fig_single_particle}
\end{figure}

In this section, we present the numerical results
for a simple case with $N=\Omega=2$.
For single-particle energies, a linear dependence on the
coordinate $q$ is assumed,
\begin{equation}
\epsilon_\alpha (q) = (-)^{\alpha+1} \epsilon(q)
                 = (-)^{\alpha+1} ( \chi q - \epsilon_0 ) \ ,
                   \quad\mbox{for } \alpha=1,2 \ .
\end{equation}
The two levels cross each other at $q=\epsilon_0 / \chi$.
A schematic figure of single-particle levels are displayed
as functions of $q$ in Fig.~\ref{Fig_single_particle}.
Each level has two-fold degeneracy.

The model can be solved quantum mechanically
using a direct product basis
$\ket{N_{\rm osc}}_{\rm core} \otimes \ket{\alpha\bar\alpha}_{\rm val}$ 
where $\ket{N_{\rm osc}}_{\rm core}$ is an eigenstate of the
core Hamiltonian with the oscillator quanta $N_{\rm osc}$
and
$\ket{\alpha\bar\alpha}_{\rm val}
\equiv c_\alpha^\dagger c_{\bar\alpha}^\dagger \ket{0}$
($\alpha=1,2$).
We use  core states with $N_{\rm osc} \leq 20$ for diagonalising
the total Hamiltonian, which can be written in a matrix form
\begin{equation}
H = \frac{1}{2}\left(p^2 + q^2 \right) I
    + \left(
     \begin{array}{cc}
      2\epsilon_1(q) - G  & -G \\
      -G  & 2\epsilon_2(q) - G
     \end{array}
     \right) \ ,
\end{equation}
where $I$ is the $2\times 2$ unit matrix.
Results obtained with this method will be referred below as
``exact calculation''.

\subsection{TDMF parametrisation}
\label{sec: results_BCS}

Now we apply the theory of ALACM
to this model with the mean-field parametrisation in
Sec.~\ref{sec: BCS_parametrization}.
The Hamiltonian has the form
\begin{eqnarray}
\Ham &=& \frac{1}{2} \left( p^2 + B_{\rm val} \pi^2 \right) + V(q,\xi) \ ,\\
\label{V_BCS}
V(q,\xi) &=& \frac{1}{2} q^2 
       + 2\sqrt{2} (\chi q - \epsilon_0 ) \xi \Xi - G (1 - \xi^2 )^2 \ ,
\end{eqnarray}
where $B_{\rm val}$ is given by Eq.(\ref{B_1dim}) with
$\epsilon = \chi q - \epsilon_0$.

\begin{figure}
\centerline{\includegraphics[width=7cm]{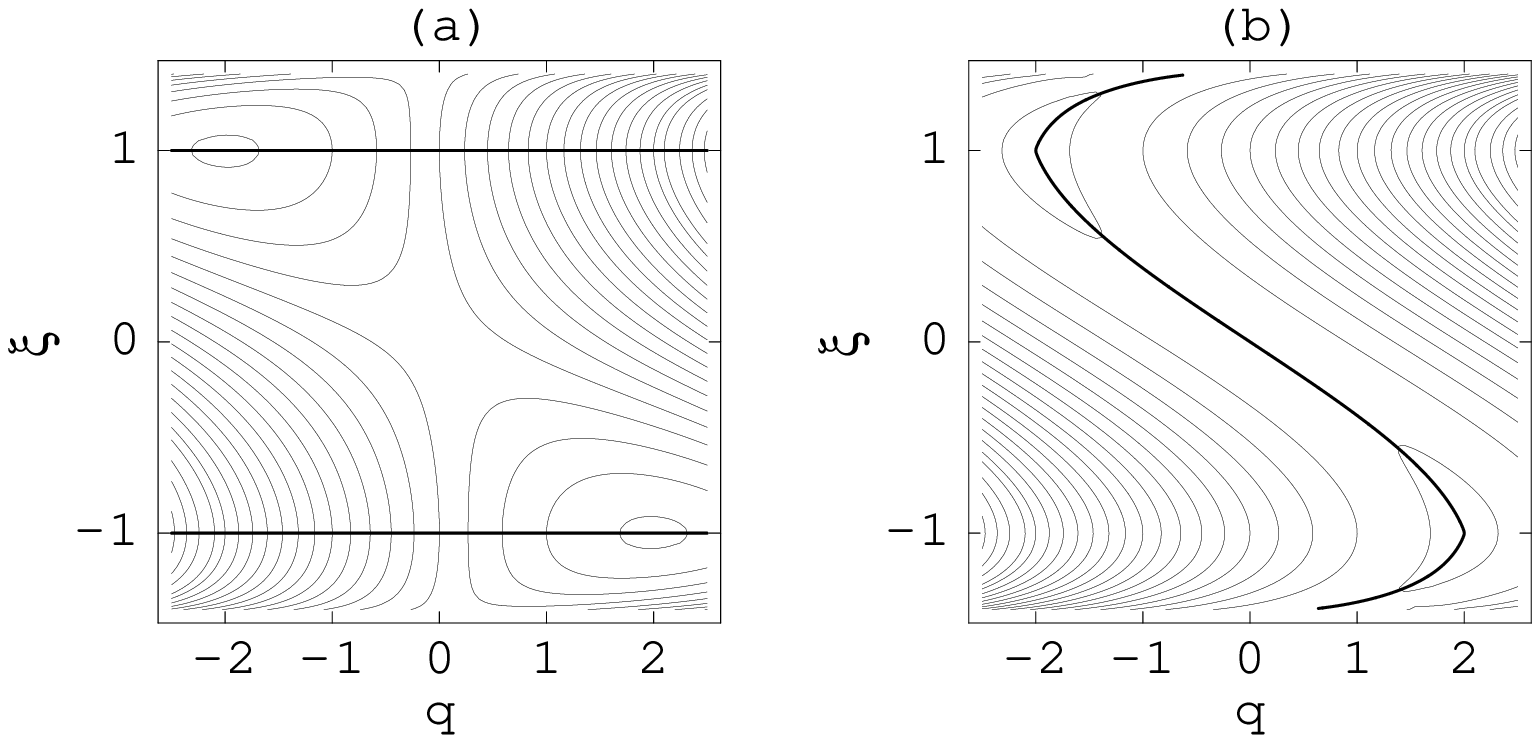}}
\caption{
Contour plots of the energy surface for the potential $V(q,\xi)$,
Eq.(\ref{V_BCS}).
The parameters $\epsilon_0=0$ and $\chi=1$ are used.
The left figure (a) is the energy surface for $G=0.2$ and
the right (b) for  $G=1.9$.
Contour lines are displayed for $\Delta V = 0.5$.
The thick lines indicate the collective paths obtained by the ALACM
theory.
}
\label{Fig_BCS}
\end{figure}

Taking $\epsilon_0=0$ and $\chi=1$,
the potential $V(q,\xi)$ has two local minima at
$(q,\xi)=(-2,1)$ and $(2,-1)$ which correspond to
valence-particle configurations $\ket{1\bar 1}$ and $\ket{2\bar 2}$,
respectively.
A collective path is now obtained by using the LHE algorithm
described in Sec.~\ref{sec: ALACM}.
We start from a local minimum, solve the RPA equations
and search for the next point in the direction of
the lowest RPA mode.
Since this system has only two degrees of freedom,
we can easily visualise the path on the 2-dimensional surface.
In Fig.~\ref{Fig_BCS}, the potential energy surface and
the obtained collective path are shown
for two strengths of the pairing force $G=0.2$ and $G=1.9$.
In the case of $\epsilon_0=0$,
the potential landscape has a symmetry
with respect to a rotation of $180^\circ$ about the origin,
$V(q,\xi) = V(-q,-\xi)$.
For a weak pairing force (Fig.~\ref{Fig_BCS} (a)),
each local minimum has an independent collective path
which represents a harmonic oscillator with a fixed valence
configuration (both particles in level 1 or level 2).
These represent diabatic solutions which do not
mix in this approximation.
On the other hand, for a strong pairing force
(Fig.~\ref{Fig_BCS} (b)),
we get a collective path which changes the particle configuration
and connects the two local minima.

We have thus found that the system automatically selects
different type of collective paths according to the strength of
pairing force.
In this approximation, we get completely diabatic solutions for
$G=0.2$ which will result in no parity splitting upon 
quantisation.
Of course, for $G\neq 0$, an exact calculation always gives
some splitting in energy.
However, the splitting turns out to be extremely small for the
weak pairing case (see below).

\begin{figure}
\centerline{\includegraphics[width=7cm]{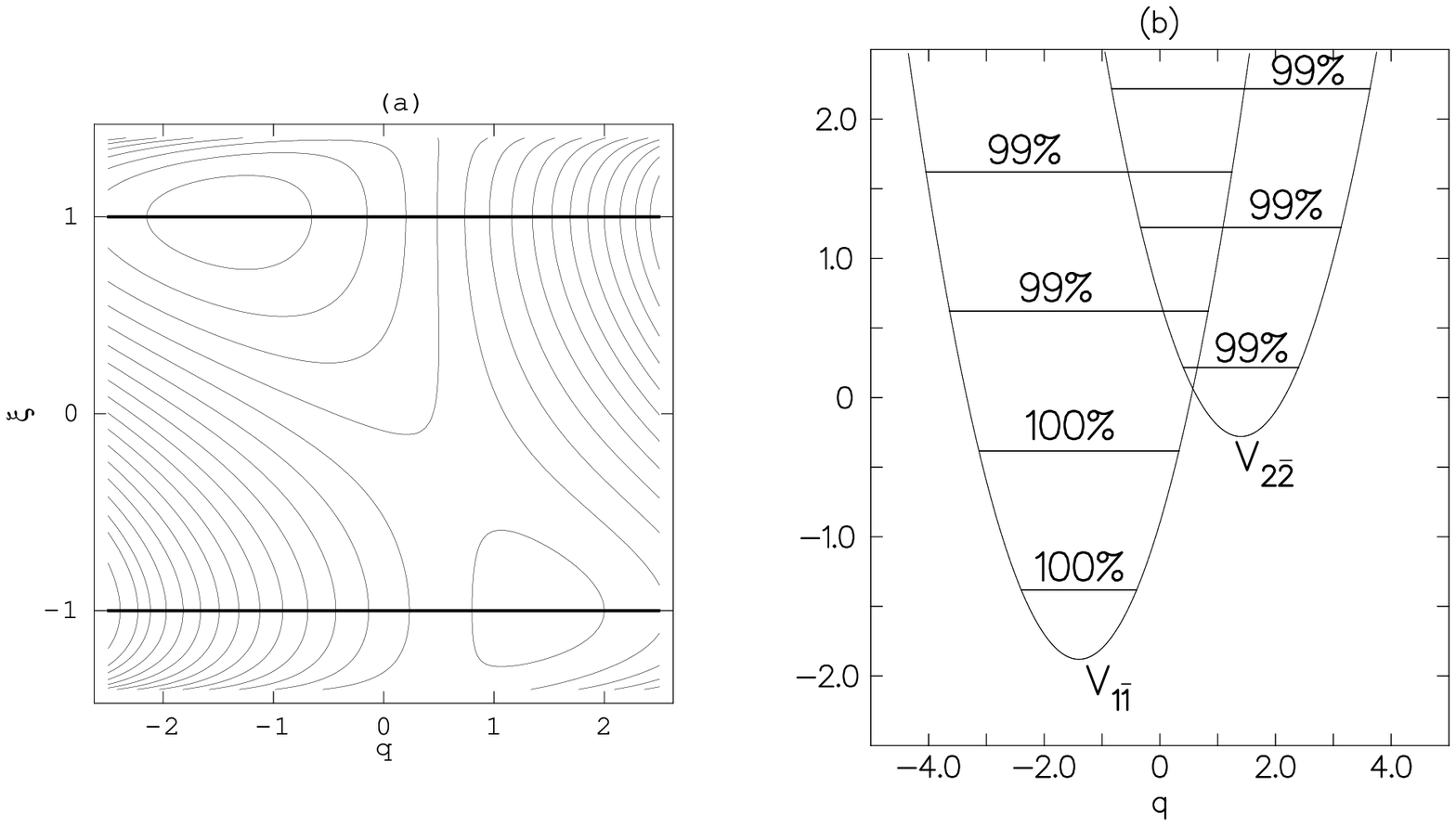}}
\caption{(a) The same as Fig.~\protect{\ref{Fig_BCS}} but for $\epsilon=0.4$,
$\chi=0.7$ and $G=0.1$.
(b) Diabatic potential curves as functions of $q$ and
eigenenergies of the exact calculation.
The number on each level indicates the percentage of the main
configuration in the wave function.}
\label{Fig_BCS_2}
\end{figure}

Next, 
in order to discuss a case of asymmetric potential landscape,
let us take $\epsilon=0.4$, $\chi=0.7$ and $G=0.1$.
Fig.~\ref{Fig_BCS_2} (a) shows the potential energy surface
and collective path.
The potential, which is no longer symmetric about the $180^\circ$
rotation about the origin,
has a deep minimum at $(q,\xi)=(-1.4,1)$ and
a shallow minimum at $q=(1.4,-1)$.
Each minimum has again an independent collective path
parallel to the horizontal line ($q$-axis),
representing the diabatic dynamics.
In Fig.~\ref{Fig_BCS_2} (b), we show results of exact calculation
for this parameter set, together with the diabatic potential
curves corresponding to the valence configuration
$\ket{1\bar 1}$ and $\ket{2\bar 2}$.
The number on each level indicates the percentage of the main
configuration in the wave function.
The exact calculation shows strong diabaticity of the system
and two different configurations hardly mix.
This agrees with the independent paths obtained in the ALACM.
Therefore, one may state that
the ALACM combined with the BCS approximation
can reproduce the qualitative feature of
the pairing systems.

\subsection{Born-Oppenheimer (BO) approximation}

Before discussing another parametrisation in the next section,
let us review the Born-Oppenheimer (BO) approximation.
The formulation in the next section can be seen as the generalised
version of BO theory,
in which a collective coordinate is chosen self-consistently.

\begin{figure}
\centerline{\includegraphics[width=7cm,angle=-90]{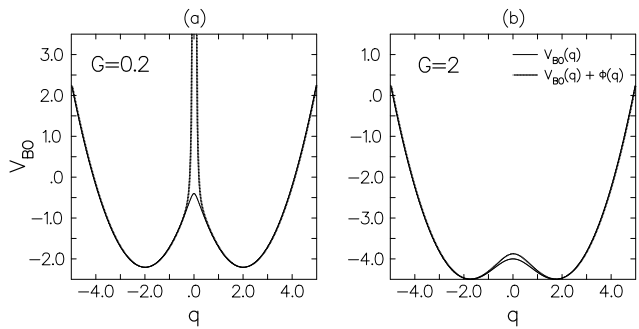}}
\caption{
Born-Oppenheimer potentials $V(q)$ (solid lines),
Eq.(\ref{V_BO}),
for $G=0.2$ (a) and for $G=2$ (b).
The parameters $\epsilon_0=0$ and $\chi=1$ are used.
The dotted lines indicate the potential including
the scalar gauge potential,  $V(q)+\Phi(q)$.
See text for details.
}
\label{Fig_BO}
\end{figure}

Using the conventional adiabatic theory which assumes motion of the
valence particles is much faster than motion of the core,
the effective Hamiltonian can be obtained by diagonalising
$H_{\rm val}$ at each value of $q$.
\begin{eqnarray}
\label{H_BO}
\bar{H}_{\rm BO} &=& \frac{p^2}{2} + V_{\rm BO}(q) \ , \\
\label{V_BO}
V_{\rm BO}(q) &=& \frac{q^2}{2}
        - \sqrt{4(\chi q - \epsilon_0)^2 + G^2} - G \ .
\end{eqnarray}
If we neglect the interaction ($G=0$),
the potential $V_{\rm BO}(q)$ has two local minima at $q=\pm \chi$
corresponding to different valence-particle configurations.
Fig.~\ref{Fig_BO} shows the BO potentials $V_{\rm BO}(q)$
for weak (Fig.~\ref{Fig_BO} (a)) and strong pairing force
(Fig.~\ref{Fig_BO} (b))
with $\chi=1$ and $\epsilon_0=0$.
In Ref.\cite{FMM91} it has been shown that
this approximation works well for the strong pairing case
but it considerably overestimates the mixing between two minima
for the weak pairing case.

One possible way to improve the BO theory is
to determine a collective path self-consistently
and take into account the effect of mass parameter $B_{\rm val}$.
This will be discussed in the next section.
The other, which is discussed here, is
to take account of the (Berry's) ``gauge'' potentials
which emerge from the derivative of the eigenstates
of the fast degrees of freedom (valence particles) $\ket{n}$
with respect to the slow coordinate $q$.
The new effective Hamiltonian will be
\begin{equation}
\tilde{H}_{\rm BO} = \frac{1}{2}\left( p - A_n(q) \right)^2 + V_{\rm BO}(q)
                        + \Phi_n(q) \ , \\
\end{equation}
where
\begin{eqnarray}
\label{A_BO}
A_n (q) &\equiv& i \inproduct{n}{\partial_q n} \ ,\\
\label{Phi_BO}
\Phi_n (q) &\equiv& \frac{1}{2} \bra{\partial_q n}
             \left( 1 - \ket{n}\bra{n} \right) \ket{\partial_q n} \ .
\end{eqnarray}
In the present case,
we take $\ket{n}$ to be the lowest eigenstate
of the valence Hamiltonian 
which can be expressed as in Eq.(\ref{state_2_level}),
with $\alpha$ given by Eq.(\ref{local_minima_2_level}).
Then, the gauge potentials are calculated as
\begin{equation}
A (q) = 0 \ , \quad
\Phi(q) = \frac{\left( \chi G \right)^2}
              {2 \left\{ 4 (\chi q - \epsilon_0)^2 + G^2 \right\}^2} \ .
\end{equation}
This shows that there is no Berry's phase for the current problem,
but does imply an additional scalar potential.
The additional potential is positive definite and has
the maximum value at the crossing point $q=\epsilon_0/\chi$.
This scalar potential is shown in Fig.~\ref{Fig_BO} by dotted lines.
The peak height at a level crossing is getting higher for weaker pairing.
This term will suppress the mixing between two minima
for the case of weak pairing, and
improve the result of quantisation (see below).

\subsection{TDSE parametrisation}
In the Sec.~\ref{sec: results_BCS},
we have shown the results using the mean-field
parametrisation of the valence Hamiltonian,
which leads to a diabatic solution in the weak pairing case.
In this section, we use the parametrisation described in
Sec.~\ref{sec: exact_parametrization}.
This conserves number of particles exactly and can be regarded
as a generalised BO theory.

The total Hamiltonian is in a form
\begin{eqnarray}
\Ham &=& \frac{1}{2} \left( p^2 + B_{\rm val}\pi^2 \right) + V(q,\xi) \ ,\\
\label{V_exact_parametrization}
V(q,\xi) &=& \frac{1}{2} q^2
     + 2(\chi q - \epsilon_0) \cos 2\xi -G \left( 1 + \sin 2\xi \right) \ ,
\end{eqnarray}
where $B_{\rm val}$ is given by Eq.(\ref{B_exact_param}).

With $\epsilon_0=0$ and $\chi=1$,
the potential surface has again two local minima at
\begin{eqnarray}
\label{local_minima}
(q,\xi) &=& \left( -\frac{\sqrt{16 - G^2}}{2},
    \frac{1}{2}\arccos\left(\frac{\sqrt{16 - G^2}}{4} \right) \right) \ ,
     \nonumber \\
 && \mbox{and }
  \left( \frac{\sqrt{16 - G^2}}{2},
    \frac{1}{2}\arccos\left(-\frac{\sqrt{16 - G^2}}{4} \right) \right) \ ,
\end{eqnarray}
and a saddle point at $(0,\pi/4)$ where we restrict $\xi$
to the region $0\leq \xi \leq \pi/2$.
In this parametrisation, $\xi$ represents the mixing angle between the two
valence configurations, $\ket{1\bar 1}$ and $\ket{2\bar 2}$.
If we minimise the potential with respect to the coordinate $\xi$
at each value of $q$,
the potential $V(q,\xi_0(q))$ becomes the same as the BO potential
(\ref{V_BO}).
If $G=0$,
the local minima correspond to the pure configurations,
$\ket{1\bar 1}$ ($\xi=0$) and $\ket{2\bar 2}$ ($\xi=\pi/2$).
In general, these minima have finite values of mixing angle,
which was not the case in the mean-field parametrisation.

\begin{figure}
\centerline{\includegraphics[width=7cm]{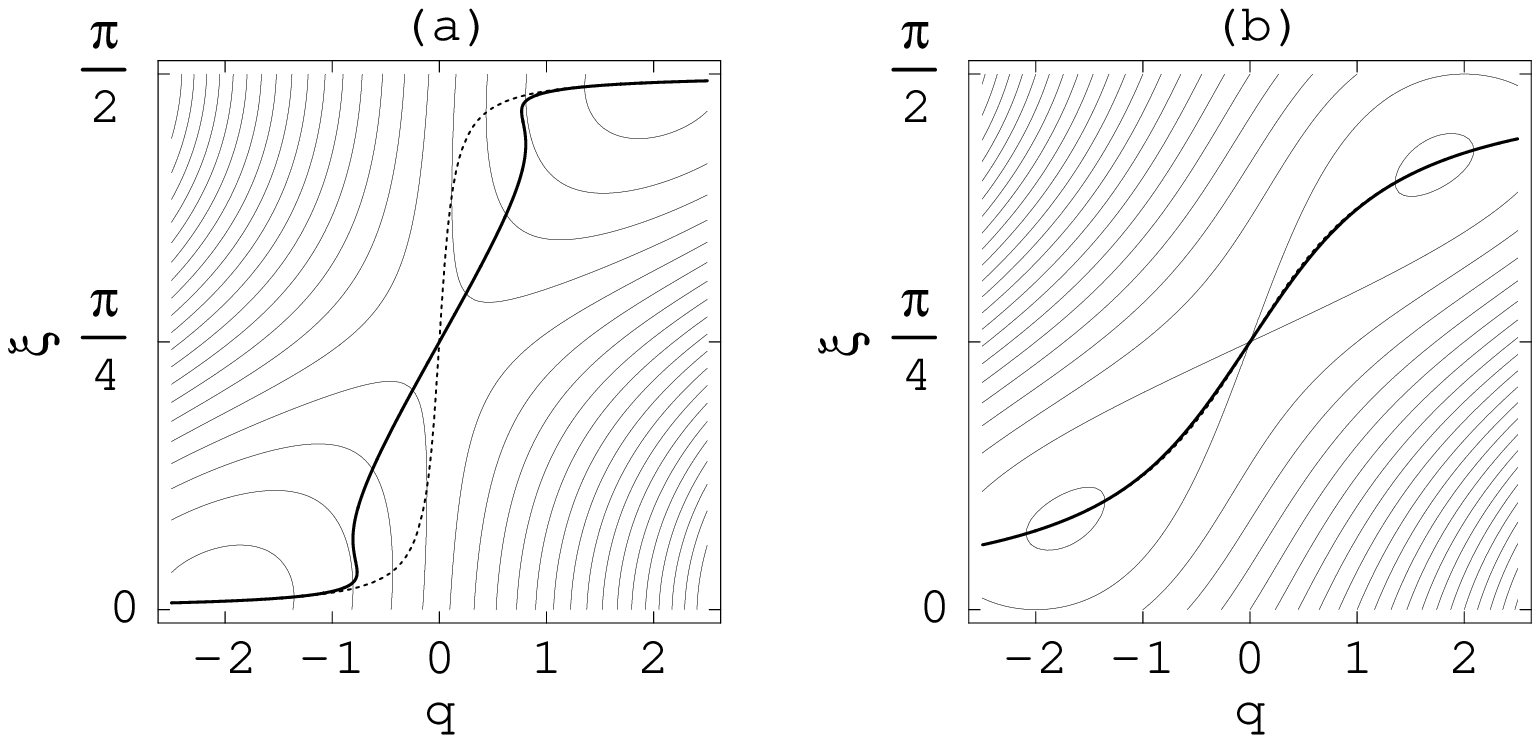}}
\caption{
Contour plots of the potential energy $V(q,\xi)$,
Eq.(\ref{V_exact_parametrization}),
for $G=0.2$ (a) and $G=2$ (b).
The parameters $\epsilon_0=0$ and $\chi=1$ are used.
The thick lines indicate the collective paths obtained by the ALACM,
while the dotted lines are the paths in the BO approximation.
}
\label{Fig_ALACM_contour}
\end{figure}

Fig.~\ref{Fig_ALACM_contour}
shows the  potential energy surfaces and the 1-dimensional
collective paths.
For a strong pairing force (Fig.~\ref{Fig_ALACM_contour} (b)),
we have a smooth path which connects two local minima.
On the other hand, for the case of weak pairing
(Fig.~\ref{Fig_ALACM_contour} (a)),
the path exhibits a peculiar kink-like
behaviour near $q=\pm 0.8$.
Close to a local minimum point, the collective path represents
a harmonic oscillator (the core Hamiltonian) which is almost parallel
to a $q$-axis (horizontal line).
On the other hand, at the saddle point $(q,\xi)=(0,\pi/4)$,
the collective path is trying to go through the valley of potential.
These paths are not able to be connected smoothly
for the weak pairing case,
which results in the strange back-bending.
In the vicinity of this back-bending,
the decoupling from the other degree of freedom is very bad
(see discussion below).

In the same figure, we show a path (dotted line)
which is obtained by minimising the potential energy
at each value of $q$ (BO approximation).
For weak pairing, the path obtained by the ALACM theory
is found to be quite different from that by the BO theory.
On the other hand,
for  strong pairing, we can hardly see the difference between them.
In the BO approximation, the motion along $\xi$-axis is assumed to be
much faster than that along $q$-axis.
In other words, the mass of valence Hamiltonian $1/B_{\rm val}$
should be much smaller than
the mass of core Hamiltonian, which can be expressed as the condition
\begin{equation}
G \gg \sin 2\xi \ .
\end{equation}
It is obvious from this equation that the BO approximation
fails for the weak pairing force $G \ll 1$.

In order to requantise the obtained collective Hamiltonian,
it is convenient to define the collective coordinate $s$ on the path
so as to make the collective mass parameter $\bar{B}^{11}=1$.
We use the relation
\begin{equation}
\label{mass_renormalization}
\left(\frac{dq}{ds}\right)^2 + \frac{\sin 2\xi}{G}
                            \left(\frac{d\xi}{ds}\right)^2
= \left(\bar B\right)^{-1} = 1 \ ,
\end{equation}
to renormalise the effect of collective mass into a new coordinate $s$.
The requantised collective Hamiltonian will be in a form
\begin{equation}
\label{ALACM_Hamiltonian}
\hat \Ham = -\frac{1}{2}\frac{d^2}{ds^2} + {\bar V}(s) \ ,
\end{equation}
where ${\bar V}(s) \equiv V(q(s),\xi(s))$.

\begin{figure}
\centerline{\includegraphics[width=7cm,angle=-90]{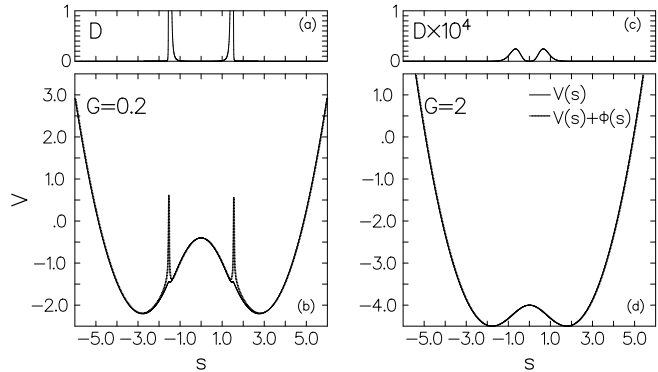}}
\caption{
Potentials along the paths (thick lines)
displayed in Fig.~\ref{Fig_ALACM_contour}
as functions of a collective coordinate
$s$ for $G=0.2$ (b) and $G=2$ (d).
Solid (dotted) lines are the potentials without (with)
a scalar gauge potential $\Phi(s)$.
The upper part, (a) and (c),
shows the decoupling measure $D$ defined
in Eq.(\ref{decoupling}) for $G=0.2$ and $G=2$, respectively.
}
\label{Fig_ALACM_potential}
\end{figure}

In Fig.~\ref{Fig_ALACM_potential},
the potential and the decoupling measure $D$ (\ref{decoupling})
are shown as functions of coordinate $s$.
Comparing these potentials with the BO potentials in Fig.~\ref{Fig_BO},
for the strong pairing case,
we see that they resemble each other and
the decoupling is found to be very good everywhere on the path.
For the weak pairing force, however,
the distance between two minima of the potential is larger
in the ALACM than in the BO theory.
The decoupling is extremely bad where the path has a kink
(the decoupling measure $D$ is order of $10^3$ at the peak),
though the decoupling is relatively good anywhere else on the path.

As we have done for the BO approximation,
we can incorporate the effective ``gauge'' potentials
into the ALACM theory.
We assume  gaussian wave functions for the intrinsic
(non-collective) degrees of
freedom (``harmonic approximation'').
Here, collective and intrinsic coordinates are expressed by $s$ and 
$x$, respectively.
The wave function is approximated in a form of semi-direct product,
\begin{equation}
\label{ALACM_Berry_wavefunction}
\Psi(s,x) = \psi_{\rm col}(s) \otimes \psi_{\rm intr}(s,x) \ ,
\end{equation}
where the wave functions for the intrinsic motion $\psi_{\rm intr}(s,x)$
are assumed to be a gaussian,
\begin{equation}
\psi_{\rm intr}(s,x) =
                 \left( \frac{\omega_x(s)}{\pi} \right)^{1/4}
                 \exp\left(-\frac{\omega_x(s)}{2} x^2 \right) \ .
\end{equation}
Here the frequency $\omega_x(s)$ are calculated by the covariant
RPA at each point on the collective path.
Since the lowest RPA mode is supposed to be
along the collective path,
$\omega_x(s)$ is the frequency of the second RPA mode.

Identifying $\psi_{\rm intr}(s,x)$ with a state $\ket{n}$ in
Eqs.(\ref{A_BO}) and (\ref{Phi_BO}),
the (Berry's) gauge potentials become
\begin{eqnarray}
\label{A_ALACM}
A(s) &\equiv& i \int_{-\infty}^\infty dx
           \psi_{\rm intr}^* \frac{\partial\psi_{\rm intr}}{\partial s}
          = 0 \ , \\
\label{Phi_ALACM}
\Phi(s) &\equiv& \frac{1}{2} \int_{-\infty}^\infty dx
           \left| \frac{\partial \psi_{\rm intr}}{\partial s} \right|^2
          = \frac{1}{16}\left( \frac{d}{ds} \ln \omega(s) \right)^2 \ .
\end{eqnarray}
Again, the vector potential $A$ vanishes and the scalar potential is
positive definite.
In Fig.~\ref{Fig_ALACM_potential},
we show this scalar gauge potential as a function 
of $s$ (dotted lines).
For a strong pairing force $\Phi(s) \approx 0$ everywhere
(one cannot see the dotted line in the figure because it  almost 
coincides with the  solid line),
however, for weak pairing,
it has sharp peaks at kinks where the decoupling
measure $D$ also has large values.
The requantised Hamiltonian (\ref{ALACM_Hamiltonian}) will be modified into
\begin{equation}
\hat\Ham = -\frac{1}{2} \frac{d^2}{ds^2} + {\bar V}(s) + \Phi(s) \ .
\end{equation}
The wave function $\psi_{\rm col}(s)$ in Eq.(\ref{ALACM_Berry_wavefunction})
is required to be an eigenfunction of this modified Hamiltonian.

\begin{figure}
\centerline{\includegraphics[width=7cm]{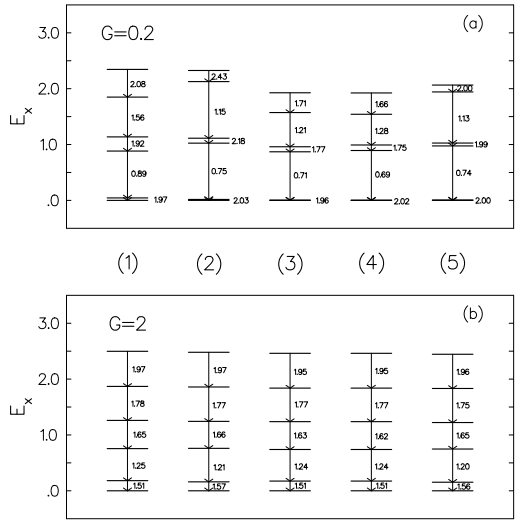}}
\caption{
Spectra up to the fifth excited states for
(1) the BO theory without $\Phi(q)$,
(2) the BO theory with $\Phi(q)$,
(3) the ALACM theory without $\Phi(s)$,
(4) the ALACM theory with $\Phi(s)$,
and (5) the exact calculation.
Figure (a) is for $G=0.2$ and (b) for $G=2$.
The numbers next to the arrows
denote The values of the transition matrix element
$|\bra{n} q \ket{n+1} |$.
}
\label{Fig_spectra}
\end{figure}

The results after requantisation are summarised in Fig.~\ref{Fig_spectra}.
There we show the spectra of (1) the BO theory without gauge potentials,
(2) the BO theory with gauge potentials,
(3) the ALACM theory without gauge potentials,
(4) the ALACM theory with gauge potentials,
and (5) the exact calculation.
For a strong pairing force, all approximation schemes work well.
On the other hand, for a weak pairing force, the result (1) overestimates
the splitting of parity doublets.
This overestimate is
corrected by including the scalar gauge potential (2).
The same effect also appears for ALACM results (3) and (4).
However, in this case,
the suppression of parity splitting originates
in the mass parameter, not in the gauge potentials.
Since we renormalise the mass parameter with
Eq.(\ref{mass_renormalization}),
this produces a larger distance between two local minima
of the potential $\bar V(s)$, as compared to $V_{\rm BO}(q)$.
The tunnelling probability between two minima
is suppressed by this mass-renormalisation effect.
The effect of gauge potentials, the difference between (3) and (4),
turns out not to be significant in the ALACM.
The agreement with the exact calculation is good
for low-lying states (up to the third excited state),
though, as expected, it is not so good for higher-lying states.

We also show
transition matrix elements $|\bra{n'} q \ket{n} |$
in the same figure.
They are calculated by
\begin{eqnarray}
\bra{n'} q \ket{n} &=&
\int_{-\infty}^\infty dq \ \psi_{n'}^{*}(q) q \psi_n(q) 
                                        \quad \mbox{for BO} \ ,\\
\bra{n'} q \ket{n} &=&
\int_{-\infty}^\infty ds \ \psi_{n'}^{*}(s) q(s) \psi_n(s)
                                        \quad \mbox{for ALACM} \ ,
\end{eqnarray}
where the wave functions $\psi(q)$ and $\psi(s)$ are normalised
with respect to the coordinates $q$ and $s$, respectively.
Again, for the case of weak pairing,
the amplitude between the first and second exited states is
overestimated in BO theory (1) due to too strong mixing of two minima.
This is corrected in ALACM (3), (4) and in BO with gauge potential (2).

\begin{table}
\caption{
Expectation values of the decoupling measure $\bra{n} D \ket{n}$
with respect to
individual eigenfunctions for cases (3) and (4)
in Fig.~\ref{Fig_spectra}.
See the main text and the caption of Fig.~\ref{Fig_spectra} for more details.
}
\begin{tabular}{c|cc|cc}
 & \multicolumn{2}{c}{$G=0.2$} &
   \multicolumn{2}{c}{$G=2$} \\
$n$ & (3)  & (4)  & (3)  & (4) \\ \hline
1 &  57  &  45   &  $3.97\times 10^{-6}$ & $3.98\times 10^{-6}$ \\
2 &  55  &  45   &  $1.72\times 10^{-6}$ & $1.71\times 10^{-6}$ \\
3 &  172 &  162  &  $3.17\times 10^{-6}$ & $3.17\times 10^{-6}$ \\
4 &  181 &  167  &  $3.19\times 10^{-6}$ & $3.19\times 10^{-6}$ \\
5 &  8.0 &  7.1  &  $1.01\times 10^{-6}$ & $1.00\times 10^{-6}$ \\
6 &  119 &  125  &  $2.99\times 10^{-6}$ & $2.99\times 10^{-6}$
\end{tabular}

\label{table_D}
\end{table}

Table~\ref{table_D} shows the expectation values of the decoupling
measure $D$ with respect to individual states,
\begin{equation}
\bra{n} D \ket{n} =
\int_{-\infty}^\infty ds \ \psi_n^{*}(s) D(s) \psi_n(s) \ .
\end{equation}
These values are about 7 orders of magnitude smaller for the case
of strong pairing $G=2$ compared to the weak pairing $G=0.2$.
For the weak pairing force,
it turns out that
all states considerably suffer bad decoupling.
Surprisingly, still we have obtained reasonable agreement for
the low-lying spectra and transition amplitudes.

\section{Conclusion}
\label{sec: conclusions}

We have investigated a simple system with level crossings
where the monopole pairing interaction plays an important role
and the potential landscape exhibits multiple
local minima.
To apply the ALACM theory,
canonical coordinates and
a classical Hamilton's equation have been introduced by
parametrisation of the time-dependent states.
We have done this in two different ways.
One is based on the time-dependent mean-field (BCS) theory
in which states are not eigenstates of particle number.
It is inevitable that this introduces a Nambu-Goldstone (NG) mode
(pairing rotation) in the theory.
We have developed the constrained local harmonic formalism which can
treat this spurious component and find a path orthogonal to the
NG mode in the configuration space.
The other is based on the exact time-dependent Schr\"odinger equation.
It is possible to derive a classical Hamilton's equation
for a simple two-level system
by regarding the mixing angle and the relative phase
as canonical variables,
though this becomes more complicated when the number of levels increases.
In this parametrisation, states are always eigenstates of particle
number and there is no NG mode.

Using the mean-field parametrisation (neglecting Fock terms),
we have found that the system exhibits
completely different collective paths
depending on the strength of pairing force.
For a weak pairing force, 
the collective path cannot connect one local minimum with another.
On the other hand, a path which connects two local minima
is obtained for the case of a strong pairing force.
In this model, different local minima correspond to different
configurations of valence particles.
Therefore, the strength of pairing force determines
either adiabatic or diabatic behaviours of the system.

Using the parametrisation of the exact time-dependent Schr\"odinger
equation,
if $G\neq 0$,
a collective path starting from a local minimum always passes through
a saddle point and reach the other minimum.
For the strong paring force, the path is very smooth and
is well decoupled from the other degree of freedom
orthogonal to the path.
However, for the weak pairing force, the path shows a peculiar
behaviour (back-bending) on the way
where the decoupling is very bad.
This is qualitatively consistent with the mean-field results.

We have requantised the 1-dimensional collective Hamiltonian 
and compared the spectra and transition matrix elements with exact
calculations.
For the strong pairing force, the agreement is excellent for
any kind of adiabatic approximation (BO, ALACM).
However, for the weak pairing,
the BO approximation fails to reproduce the qualitative features of
exact spectra,
while the ALACM well reproduces the low-lying spectra and transition
amplitudes.
The agreement becomes worse for the higher-lying states
beyond a barrier height between two local minima.
The theory is based on the adiabatic assumption in which
Hamiltonians are expanded with respect to momentum up to second order.
Thus, this disagreement may come from the higher-order terms
we have neglected,
because the higher-lying states are supposed to have larger momenta.

We have also discussed the effect of gauge potentials arising from
change of intrinsic states along the path.
Inclusion of these potentials turns out to improve
the BO calculations considerably
but to leave the ALACM results almost unchanged.

The authors of Ref.\cite{FMM91} have studied 
the same model Hamiltonian and concluded that
for the case of a weak pairing force
the adiabatic approach (BO approximation)
is not appropriate for describing the system
and the diabatic approach is needed.
However, one should bear in mind that,
in the BO approximation,
the collective coordinate $q$
was chosen by assuming that the motion of valence particles
described by $(\xi,\pi)$ is much faster than the motion
along $q$.
This is apparently not the case
for the weak paring $G\ll 1$.
In our adiabatic approach,
all degrees of freedom are treated equivalently,
taking account of the mass tensor in the multi-dimensional
configuration space.
Thus the  collective path is very much
different from the one in the BO approximation
for the case of weak pairing.
Actually, the collective path turns out to represent the completely
diabatic dynamics in some cases.
Therefore, if one has a decoupled path determined by the system itself,
the {\it adiabatic} theory may account for the {\it diabatic}
dynamics.

For the practical applications of the ALACM theory,
generally speaking,
it is not easy to solve the constraints about the NG mode explicitly.
Since nuclei show many kinds of symmetry breaking (translational symmetry,
rotational symmetry, gauge symmetry),
the constrained local harmonic formalism discussed in Sec.~\ref{sec: CLHE}
will be very useful in such cases.
We shall apply this method to more realistic problems in the near future.

In conclusion, the ALACM improves on the conventional BO theory.
It works well even when the mass of valence Hamiltonian is smaller
than that of the core Hamiltonian,
because, in principle, the ALACM treats all degrees of freedom
equivalently.
The ALACM fails to reproduce properties of higher-lying states
in the case of weak pairing.
This is definitely due to a limitation of the adiabatic theory.
Upon  requantisation, a reasonable agreement with the exact
calculation has been achieved in both cases of strong and weak
pairing forces, at least for the low-lying states.

\acknowledgements

This work was supported by a research grant from the Engineering
and Physical Sciences Research Council (EPSRC)  of Great Britain.

\end{document}